\documentclass{SCGE}
\usepackage{graphicx}
\usepackage{rotating}
\begin{document}

%%%%%%新版式要加上这组
\begin{picture}(0,0){\rm
\put(0,-20){\makebox[160truemm][l]{\bf {\sanhao\raisebox{2pt}{.}}
News and Views  {\sanhao\raisebox{1.5pt}{.}}}}}
%\put(0,-34){\jiuwuhao {\textcolor[rgb]{0.5,0.5,0.5}{\sf Progress of Projects Supported by %NSFC}}}%%(11??×￠êí￡oμ÷\textcolor[rgb]{x,x,x}?Dμ?êy×?x??′ó???ò)
\end{picture}

\def\bm{\boldsymbol}

\def\dl{\displaystyle}
\def\du{\end{document}}
\def\d{{\rm d}}
\def\e{{\rm e}}
\def\i{{\rm i}}

\def\pi{{\uppi}}

% The author doesn't need fill in it.
\Year{2018} %
\Month{October} %
\Vol{61} %  ?ío?
\No{10} %  ?úo?
\BeginPage{1} % ?eò3??
\AuthorMark{{\rm Xu}}  %(11??×￠êí￡oò3??é?μ?×÷??)
\AuthorMarkCite{{\rm Xu}. } %(11??×￠êí￡ocitation?Dμ?×÷??)
\DOI{https://doi.org/10.1007/s11433-018-9217-y} % The author doesn't need fill in it.
\ArtNo{109531}

% \title[short text for running head]{full title}{comments for title}
\title[Strong Matter]{Strong Matter: Rethinking Philosophically}

\author[]{Renxin XU}{}

\address[]{School of Physics and KIAA (Kavli Institute for Astronomy and Astrophysics), Peking University, Beijing 100871, China; r.x.xu@pku.edu.cn}
%\address[{\rm2}]{Center of Theoretical Nuclear Physics, National Laboratory of Heavy-Ion Accelerator, Lanzhou 730000, China}

\maketitle \vspace{-3.5mm}{\footnotesize\begin{center}
Received February 13, 2018; accepted March 7, 2018; published online May 20, 2018
\end{center}}\vspace*{-5mm}

%     Abstract is required.
%\begin{center}
%\rule{16.5cm}{0.4pt}
%\parbox{16.5cm}
%{\begin{abstract}
%Pulsars are good clocks in the universe. One fundamental question is that why they are good %\end{abstract}}
%\end{center}\vspace*{-0.6cm}

%\begin{center}
%\parbox{16.5cm}
%{\bf\jiuhao magnetar, magnetic field, neutron star, pulsar, wind}%1??ü′ê
%\end{center}

\begin{center}
%{\PACS{\rm 23.40.-s, 23.40.Bw}}%・?àào?
%\CITA    %%(11??×￠êí￡oCitation?úèY×??ˉéú3é)
%
{\color{blue} \em Normal condensed matter is merely of electromagnetic interaction. %
A novel state of strong-interaction matter is revisited.}\\
\Cit{R. X. Xu, Strong Matter: Rethinking Philosophically, Sci. China-Phys. Mech. Astron. {\bf 61}, 109531 (2018), https://doi.org/10.1007/s11433-018-9217-y}%%(11??×￠êí￡oCitation ?úèYDèê??ˉì?D′)
\end{center}

\textwidth=178truemm \textheight=236truemm%%%%%%D?°?ê?òa?óé?

%%%%%%%%%%%%%%%%%%%%%%%%%%%%%%%%%%%%%%%%%%%%%%%%%%%%%%%%%%%%
\wuhao\vspace*{1.5mm}

\begin{multicols}{2}

%%%%%%%%%%%%%%%%%%%%%%%%%%%%%%%%%%%%%%%%%%%%%%%%%%%%%%%%%%%%
%% Text of article.
%%%%%%%%%%%%%%%%%%%%%%%%%%%%%%%%%%%%%%%%%%%%%%%%%%%%%%%%%%%%
%    Section headings
\renewcommand{\baselinestretch}{1.08} \baselineskip 12.2pt\parindent=10.8pt

\renewcommand{\thefootnote}

%\noindent
%Dear Editors,

%\vspace{2mm}
\noindent
As we continue to comprehend the underlying nature of matter in different forms, human civilization develops further.
Normal matter in everyday life, the basic stuff of which
was posited by speculators even before Socrates, is dominated
by electromagnetic interaction, and it could hence be simply
termed {\it electric matter}. Analogously, gravitationally controlled
systems (e.g., galaxies and galaxy clusters) belong to {\it
gravitational matter}, and atomic nuclei to {\it strong matter}.
What if normal baryonic matter is compressed intensely by gravity so that a huge number of nuclei would gradually merge to form a gigantic nucleus?
This is certainly of strong matter, and will be discussed succinctly here.
We argue that 2-flavored (u and d quarks) nucleons constitute
a microscopic nucleus, but 3-flavored (u, d and s quarks) strangeons
constitute a gigantic nucleus.
This has profound implications in astrophysics and cosmology, as explained in this essay.

Rational thinking about strong matter in bulk dates back to 1920s when
Rutherford's atomic nucleus model was popularized,
and a breakthrough occurred with the discovery of radio pulsars in 1967;
however, the real nature of pulsar is essentially related to the
fundamental strong interaction at low-energy scale and hence to the
non-perturbative quantum chromo-dynamics (QCD), which still remains a challenge.
Nevertheless, it is generally thought that strangeness would play
an important role in understanding the state of bulk matter at
supra-nuclear density, and that the unknown state could be the first
big problem to be solved in the era of gravitational-wave astronomy.

{\em The energy scale and its impacts}.
For strong matter at a few nuclear densities, the separation between quarks is $\sim 0.5$ fm, and the energy scale is thus of order $E_{\rm scale}\sim 0.5$ GeV, according to Heisenberg's relation.
The perturbative QCD, based on asymptotic freedom, works well at energy scale of $\Lambda_\chi>1$ GeV, whereas the mass difference between strange and up/down quarks is $\Delta m_{\rm uds}\sim 0.1$ GeV; therefore, $\Delta m_{\rm uds}\ll E_{\rm scale}<\Lambda_\chi$.
This fact has three impacts on the nature of strong matter.
(1). Chiral symmetry would be broken and quarks would be dressed with mass $\tilde{m}_{\rm q}\sim 0.3$ GeV, as is evident from both lattice-QCD and approaches of Dyson-Schwinger equations.
(2). The coupling could still be strong, even with constant $\alpha_{\rm s}\gtrsim 1$.
We can estimate the typical density of strong matter in a way similar to that of electric matter.
For the simplest electricity-bound system (i.e., hydrogen atom) with length $l_{\rm ep}$ and interaction energy $E_{\rm ep}$, equating kinematic and potential energies would result to
\begin{equation}
l_{\rm ep} \sim {\hbar^2/(m_{\rm e}e^2)} = {1\over \alpha_{\rm em}} {\hbar
c\over m_{\rm e}c^2},\; \; E_{\rm ep} \sim \alpha_{\rm em}{\hbar c\over
l_{\rm ep}} = \alpha_{\rm em}^2m_{\rm e}c^2,
\label{atom}%
\end{equation}
where $\alpha_{\rm em}=e^2/(\hbar c)\simeq 1/137$ is the coupling constant
of the electromagnetic interaction.
So the typical density reads,
\begin{equation}
\rho_{\rm EM}\simeq {m_{\rm p}\over l_{\rm ep}^3} =
({\alpha_{\rm em}c \over \hbar})^3 m_{\rm e}^3m_{\rm p} \sim 10~{\rm g/cm^3},
\label{rho_EM}%
\end{equation}
where $m_{\rm e}$ and $m_{\rm p}$ are the electron and proton masses, respectively.
Certainly the density of strong matter is relevant to the hard core between nucleons, a consequence of non-perturbative QCD.
Nevertheless, Coulomb-like interaction could occur at small distances since both gauge bosons (photon and gluon) are massless. Therefore, the density of strong matter becomes
\begin{equation}
\rho_{\rm SM}\simeq({\alpha_{\rm s}c \over \hbar})^3 \tilde{m}_{\rm q}^4 \sim 2\times 10^{15}~{\rm g/cm^3},
\label{rho_SM}%
\end{equation}
where we choose $\alpha_{\rm s}=1$ and $\tilde{m}_{\rm q}=0.3$ GeV for estimation.
Evidently, the value of $\rho_{\rm SM}$ is representative of the nuclear density, indicating that around the nuclear density, quarks would always be clustered or localized, as in the case of cold electric matter at zero pressure, rather than free.
(3). Strangeness should be included to reveal the secret, which has already been noted since 1970s~\cite{Bodmer1971,Witten1984}.
However, it has always been wondered why the stable nuclei in the world are 2-flavored. We may provide a short answer at first: nuclei are too small to have a 3-flavor symmetry, but this does not apply to huge strong matter!
The Fermi energy of electrons is negligible for micronuclei but is significant for a gigantic-nucleus produced in the core of a massive star during supernova.
Conventionally, neutronization has been the explanation for the removal of energetic electrons even since Landau~\cite{Landau1932}, but an alternative explanation could be strangenization, as further explained below.

{\em Strangeon: from nuclear symmetry energy to a principle of flavor maximization}.
Adopting a phenomological approach, nuclear physicists introduce a symmetry energy representing
the symmetry between the proton and neutron of a stable nucleus, which is essentially the balance of two flavors of quarks u and d, but the underlying physics is yet to be well understood.
In nuclear Fermi gas model, the kinetic term of nuclear symmetry energy has the same isospin asymmetry dependence; however the strong interaction is surely not negligible in reality and a potential term could dominate.
Compelling evidence for the potential term comes from scattering experiments, which show that the neutron-proton pairs are nearly twenty times as prevalent as proton-proton or neutron-neutron pairs~\cite{Subedi2008,Hen2014} because of short-range interactions.
In addition, while deuterium is stable, we didn't observe a nuclide-like bound state of two protons ($^2$He), even though the neutron mass, $m_{\rm n}$, is higher than that of proton in vacuum, with a mass difference $(m_{\rm n}-m_{\rm p})$ of 0.8 MeV, which is comparable to, if not higher than the possible Coulomb energy of that state.
These hint that a balance between quark flavors may play a key role for a stable state of strong matter.

\begin{figure}[H]
\begin{center}
%\hspace{-5cm}
%\begin{rotate}{270}
\includegraphics[width=0.22\textwidth]{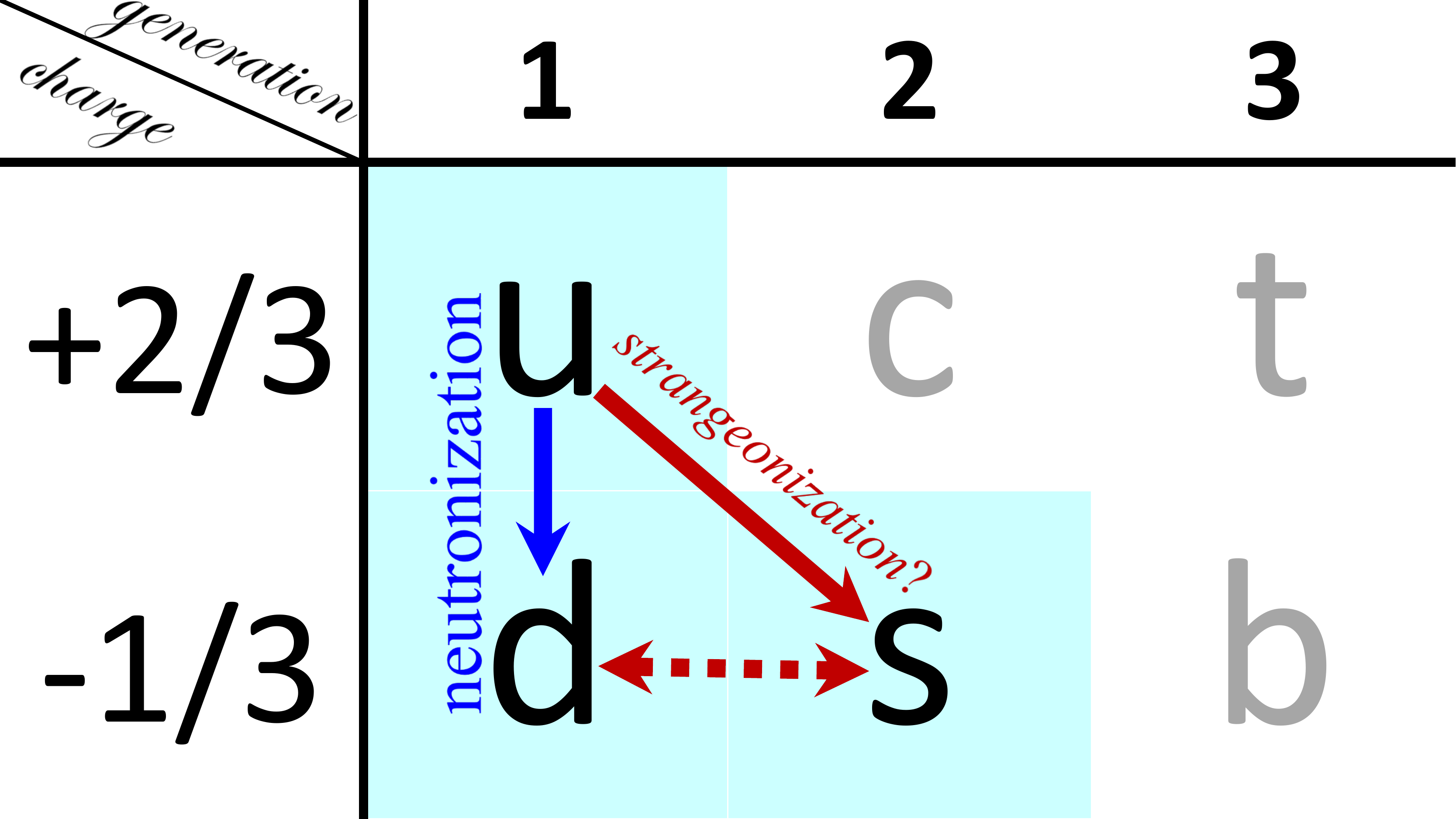}
%\end{rotate}
\hfill
\end{center}
%\vspace{-1cm}
\caption{There are six flavors of quarks in the standard model of particle physics, but only the light flavors (u, d, and s) work for strong matter. Normal nuclei are 2-flavored since the accompanied electrons are non-relativistic and the mass difference between strange and up/down quarks is much higher than the electron rest mass, while gigantic nuclei could be 3-flavored because of both flavor maximization and charge neutrality with negligible electrons.
}
\label{fig}
\end{figure}
Electrons are indeed needed to maintain electric neutrality for 2-flavored strong matter, but are not necessary for 3-flavored one (Fig.\ref{fig}).
The electron probability density inside a nucleus is negligible because of its smallness ($\ll$ the electron Compton wavelength, $\lambda_{\rm e}\simeq 10^3$ fm); hence electrons are non-relativistic. Nonetheless, they possess relativistic energy for strong matter with baryon number of $A > (\lambda_{\rm e}/{\rm fm})^3\sim 10^9$, provided a 2-flavor symmetry is maintained.
For a huge nucleus with $A > 10^9$, things would change if 3-flavor symmetry is restored.
This possibility might be real if Nature further makes use of a flavor maximization principle: {\em the short-range interaction in strong matter prefers to maximize quark flavors in order to minimize the total energy}.
The common material of the world today is 2-flavored because $\Delta m_{\rm uds}\gg m_{\rm e}$, but involving strangeness would not only maximize the flavors but also kill energetic electrons to reach a more stable state of gigantic strong matter (Fig.\ref{fig}).
Flavor maximization might be one of the working languages Nature uses, which can aid understanding of nuclear symmetry energy and help determine the possibility of strangeon matter in bulk.
In this sense, 3-flavor symmetry would be restored if strong matter is very big, although stable small nucleus would be 2-flavored.

Though there are six flavors in the standard model~(Fig.\ref{fig}), only light flavors of quarks may work for strong matter, because $m_{\rm heavy}\gg E_{\rm scale}$, where $m_{\rm heavy}$ is the current mass of heavy quark.
The absence of heavy flavors of quarks in gravity-free strong matter could also be
as a result of $m_{\rm heavy}\gtrsim\Lambda_\chi$.
Quarks would be non-local if the strong matter density is high enough to excite heavy quarks, and the state could then be softer than 3-flavored state, which would inevitably lead to the collapse of a compact star into a black hole.

Witten~\cite{Witten1984} conjectured that strange quark matter (composed of free u, d and s quarks) in bulk constitutes the true ground state of strong-interaction matter rather than $^{56}$Fe.
Bulk strangeon matter would be more stable if quarks are still clustered~\cite{Xu2003} as the color coupling can still remain significant at typical energy of $E_{\rm scale}<\Lambda_\chi$.
Therefore, during a core-collapse supernova, strangeonization, rather than neutronization, could be the cause of de-electronization ~(Fig.\ref{fig}).

{\em Test of strangeon}.
Upon different manifestations of pulsar-like objects, we propose a quark-cluster state for 3-flavored strong matter~\cite{Xu2003}.
For simplicity, the term “strange quark-cluster” is renamed to {\em strangeon}, coined from “strange nucleon”.
In astrophysics, strangeon star behaves very differently from normal neutron star and strange quark star, as is summarized in \cite{LaiXu2017}.

It is worth noting that, for a strangeon star, the conjectured principle of flavor maximization would result not only in its solid state (inner structure) but also in its self-bound nature (surface condition).
As argued in previous paragraphs, we may expect that the maximum number of quark flavor in nucleus and bulk strong matter would be {\bf 2} and {\bf 3} (i.e., with strangeness), respectively.
On one hand, strangeon is more massive than nucleon and would behave like a classical particle because of its small quantum wavelength. Thus, at low temperature, strangeon matter could be solid~\cite{Xu2003} if the kinematic thermal energy is much lower than the interaction energy between strangeons.
On the other hand, normal neutron star should be gravity-bound because extremely asymmetric isospin matter is unstable on surface, but 3-flavored strangeon matter could be stable at zero pressure (i.e., self-bound) as in the case of 2-flavored nucleon matter, the nucleus.

This strangeon star model of compact object passed the two tests below, both of which are based on dynamical and model-independent measurements.
(1). Because strangeons are massive (thus non-relativistic) and could also have a hard core at short distance, the equation of state of strangeon matter would be quite stiff and the maximum mass of strangeon star could be as high as $3M_\odot$~\cite{LaiXu2009}, while causality condition is always satisfied~\cite{Lu2018}.
This stiff property is supported by later discoveries of massive pulsars (PSR J1614-2230 and PSR J0348+0432), and a pulsar with much higher mass (e.g., $\simeq 2.5M_\odot$) is still expected.
(2). Fortunately, the strangeon star model survives the scrutiny of GW 170817 for the measurement of dimensionless tidal polarizability, $\Lambda$ ~\cite{Lai2018}.
It was found that $\Lambda(\sim 1.4M_\odot)<10^3$ from the gravitational-wave observation of GW 170817~\cite{GW}; thus some models of relatively soft states, where the hyperon puzzle is unavoidable, could be favored in the regime of normal neutron star.
But for strangeon stars, $\Lambda(\sim 1.4M_\odot)\simeq 400$~\cite{Lai2018}.
It is worth noting that both hyperon degree and quark de-confinement are irrelevant to the case of strangeon matter.

Besides the model-independent tests above, the strangeon star model is useful and necessary to model a variety of observational phenomena.
Although in principle, different manifestations of pulsar-like stars could be understood with strangeon star ~\cite{LaiXu2017}, strangeon star is popularly thought as strange, like its name.
For example, after multi-messenger observations of GW 170817, one would conclude that as neutron stars are made almost entirely of neutrons as their name implies, the discovery of kilonova of GRB 170817A suggests a neutron-rich environment, and any kind of strange star model should be ruled out.
This conclusion is actually model-dependent because other mechanisms (quark-kilonova and strangeon-kilonova) can also reproduce the observations~\cite{Lai2018}, though the neutron-kilonova model seems well developed.
Strangeon matter in bulk could be more stable than nuclear matter, but the strangeon nuggets ejected during a binary strangeon star merger could be unstable under the strong and weak interactions if their baryon number $A<A_{\rm c}\sim 10^9$.
These unstable nuggets would decay into proton/neutron and maybe heavy nuclide, but certainly, substantial microphysical investigation is needed to obtain a suitable value of the critical baryon number, $A_{\rm c}$.

{\em More consequences: trinity of strangeon matter?}
Noting the importance of strangeness, Witten~\cite{Witten1984} conjectured an absolutely stable state of quark matter, realizing that the Fermi momentum could be so high that converting nonstrange quarks to strange ones might be energetically favored, and discussed dramatic consequences of strange quark matter: quark star produced during supernova, quark nuggets residual after cosmic QCD phase-transition, as well as strange cosmic ray.
These three are retained if quark flavor maximization alternatively results in a stable strong matter, and strangeon matter may share a similar trinity of compact star, dark matter, and cosmic ray.

Invisible strange quark nuggets would remain if the cosmic QCD phase transition is of first order, and dark matter could be understood as a QCD effect~\cite{Witten1984}, but quark nuggets would not survive due to boiling and evaporation.
Later research shows that the order of transition could be higher even of crossover.
However, for a strangeon nugget in which the color coupling is much stronger, there is still room for a first order phase transition if rich non-perturbative effects are included, and furthermore, the contribution of boiling and evaporation could be relatively insignificant.
In this sense, strangeon nuggets might be a safe candidate for dark matter.
The number density of strangeon dark matter around the Earth is $n_{\rm sdm}\simeq 0.1/A$ cm$^{-3}$, and it is evident that current direct detection of strangeon dark matter would be unlikely for large baryon number of strangeon nugget, where $A\gg A_{\rm c}$ (Note: $n_{\rm sdm}\simeq 10^{-16}/A_{30}$ km$^{-3}$ is $\sim 10^{-16}$ km$^{-3}$ for $A_{30}\equiv A/10^{30}=1$, while the baryon number inside a Hubble volume was of order $A\sim 10^{48}$ during the epoch of cosmic hadronization. For strangeon dark matter with dynamical velocity of $\sim 200$ km/s near Sun, the event rate of penetrating Earth and Moon are $\sim 80/A_{30}$ yr$^{-1}$ and $\sim 6/A_{30}$ yr$^{-1}$, respectively, but $\sim 10^6/A_{30}$ yr$^{-1}$ for the Sun. It would be interesting to know what happens during the penetration events, i.e., the bombardment of strangeon dark matter with mass $\sim 10^3A_{30}$ kg and radius$\sim 10 A_{30}^{1/3}~\mu$m).
Recent negative result of PandaX-II running in the China Jinping Underground
Laboratory (for an introduction to PandaX experiments, e.g., see~\cite{Panda}) provides serious challenge to some of the popular dark matter models~\cite{JiXD2017}.
One has to think about new insights into dark matter if current detectors with increasing sensitivity would further fail to catch the elusive dark matter particle, which is assumed to be extremely light.
Nevertheless, non-relativistic strangeon dark matter matters in astrophysics. For instance, a collision between gas and strangeon dark matter nugget in primordial halo would lead to formation of seed black holes, and the supermassive black holes could then be created at redshift as high as $z\sim 6$~\cite{Lai2010}. More cosmological consequences of strangeon dark matter is surely worth investigating.

Another consequence is related to cosmic ray.
Stable strangeon nuggets with baryon number $A\gtrsim A_{\rm c}$ could be ejected relativistically/non-relativistically after a merge of binary strangeon stars, and may reach Earth through long-time travel.
For a nugget with $A\simeq 10^{10}$, the rest mass is $\sim 10^{19}$ eV, and the deposit energy during corresponding air shower could then be of order $10^{18\sim 20}$ eV, depending on its Lorentz factor.
The muon excess~\cite{excess2016} detected in ultrahigh energy cosmic ray air showers may be a hint of massive strangeon nugget, and a microphysical foundation of the strangeon air-shower is welcome to model this cascade process.
Negative results of strange cosmic ray would not rule out this possibility, for the PAMELA experiment~\cite{PAMELA2015} is only sensitive to strangeon nuggets with $A<10^5$.
Supposing that strangeon cosmic ray travels freely across the intergalactic medium and has a mass density of $\sim 10^{-19}$ GeV/cm$^3$ (note: the density of normal luminous matter is $\sim 10^{-7}$ GeV/cm$^3$), one can then estimate the event rate of such energetic strangeon air-shower, $\sim 0.1A_{10}^{-1}$ km$^{-2}$ yr$^{-1}$ ($A_{10}=A/10^{10}$), to be smaller than but comparable to the detected rate of ultrahigh energy ($>5\times 10^{18}$ eV) cosmic ray.
Detailed calculations of population synthesis would be necessary to test the strangeon cosmic ray scenario.

{\em What if strangeon matter does not exist in reality?}
As discussed above, the strangeon idea seems natural, and it might be a pity if Nature failed to make use of this possibility.
The non-existence of strangeon matter would have two implications.
(1). The nuclear symmetry energy would be irrelevant to quark flavor maximization.
(2). It could be easier to change flavor in the same generation (e.g., u and d quarks) so that neutronization (neutron star consequently) works, but might somehow be difficult among three flavors (u, d and s) and then strangeonization is unfeasible.
Both these points may have theoretical implication for non-perturbative QCD, and would help in understanding the nature of strong interaction at low but natural energy scale.

{\em Summary.} The standard model (SM) of particle physics combined with Einstein's general relativity (GR), passing tests successfully, remains a firm foundation for us to unlock the cosmic secret, especially after the discoveries of Higgs boson and gravitational-wave radiation from either black hole binaries or binary neutron stars, but it does not apply to dark matter and dark energy.
However, these dark sectors could be explained by assuming a nonzero/positive cosmological constant and by applying the principle of flavor maximization, though the underlying physics remains to be investigated.
A conceptual theme, the strangeness, is rethought philosophically in this paper, which could be very necessary to understand different manifestations of strong matter in our world (from atomic nucleus to pulsar-like object, dark matter, and even cosmic ray).
In this sense, the SU(3) flavor symmetry is not only the knocking brick of sub-hadron world, but also a key to understand strong matter in the present universe.

After a relatively quick development of fundamental laws, the pace of foundational increase in physics might be slow across the next hundreds of years, as in the case of the geocentric era of Aristotle, because presently, it is still far away to reach a scale of $\sim 10^{16}$ TeV for new physics, in which SM and GR would be unified.
It is fantastic that Einstein's GR seems always right~\cite{Shao} despite the development of alternative gravity, and it is possible that SM and GR may govern for an unimaginably long period.
If so, the unknown state of pulsar-like object could be the first big problem to be completely solved with gravitational-wave astronomy; otherwise, we might have a sad future because the problem of state equation of supra-nuclear matter would be coupled inconceivably with the puzzle of alternative gravity.

In the era of multi-messenger astronomy, advanced Chinese facilities may provide unique opportunities for scientists over the world to understand the equation of state of supra-nuclear matter, a problem relevant to non-perturbative QCD.
There are some Chinese projects to detect cosmic rays, neutrinos, and gravitational waves, which could be informative, but now the most accessible and useful tools are employed across the electromagnetic spectra.
The biggest single-dish radio telescope, FAST (Five-hundred-meter Aperture Spherical radio Telescope), has just achieved its ``first-light'', and a survey plan called CRAFTS (Commensal Radio Astronomy FasT Survey~\cite{FAST}) has been made.
It could be a revolutionary event if FAST, because of its high sensitivity, would discover a pulsar as massive as $2.3M_\odot$ (note: GW170817-favored neutron stars can only have a maximum mass of $\sim 2.2M_\odot$), while higher signal-to-noise-ratio pulse profiles could also tell us about the physics of magnetospheric activity~\cite{Tong} that depends on the surface and thus on the nature of pulsar.
It is also expected to detect fast spinning radio or X-ray pulsars, with periods around or even smaller than 1 ms, that could be a clear evidence for strangeon star because its rigidity and condensed surface may result in a super-Keplerian rotation speed.
The spin frequency of a solid strangeon star should not be limited by {\em r}-mode instability, and its self-bound surface allows a much higher frequency to break up and a smaller stellar mass even as low as $< 0.1M_{\odot}$, so that the braking action torqued by its gravitational-wave emission is negligible.
The next generation X-ray telescope, enhanced X-ray Timing and Polarimetry (eXTP), to be launched around 2025, may also provide very tight statistical constraints on the dense matter equation of state~\cite{eXTP}.
Certainly, Chinese optical telescopes, both terrestrial and in space, would also contribute to solving the big problem of dense matter inside compact star.

%\vspace*{1mm}
\noindent
{\small \em The author would like to thank those involved in the continuous discussions in the pulsar group at Peking University. This work is supported by the National Key R\&D Program of China (No. 2017YFA0402602), the National Natural Science Foundation of China (Grant Nos. 11673002 and U1531243), and the Strategic Priority Research Program of CAS (No. XDB23010200).}

\end{multicols}

\end{document}